\documentclass[showpacs,preprintnumbers,pre]{revtex4}
\usepackage{bm}
\usepackage{epsf}
\begin{document}

\title{Surface Scaling Analysis of a Frustrated Spring-network Model for
Surfactant-templated Hydrogels}
\author{G.~M.\ Buend{\'\i}a$^{1,2}$}\email{buendia@usb.ve}
\author{S.~J.\ Mitchell$^{1}$}\email{s.mitchell@tue.nl}
\author{P.~A.\ Rikvold$^{1}$}\email{rikvold@csit.fsu.edu}
\affiliation{
$^1$Center for Materials Research and Technology,
School of Computational Science and Information Technology,
and Department of Physics,
Florida State University, Tallahassee, Florida 32306-4350\\
$^2$Department of Physics, Universidad
Sim{\'o}n Bol{\'\i}var, Caracas 1080, Venezuela
}

\date{\today}

\begin{abstract}
We propose and study a simplified model for the surface and bulk structures of 
crosslinked polymer gels, into which voids are introduced through templating by 
surfactant micelles. Such systems 
were recently studied by Atomic Force Microscopy 
[M.~Chakrapani {\em et al.\/}, e-print cond-mat/0112255]. 
The gel is represented by a frustrated, 
triangular network of nodes connected by springs of random equilibrium lengths. 
The nodes represent crosslinkers, and the springs correspond to 
polymer chains. The boundaries are fixed at the 
bottom, free at the top, and periodic in the lateral direction. 
Voids are introduced by deleting a proportion of the nodes and 
their associated springs. 
The model is numerically relaxed to a representative local energy minimum, 
resulting in an inhomogeneous, ``clumpy'' bulk structure. The free top 
surface is defined at evenly spaced points in the lateral ($x$) direction 
by the height of the topmost spring, measured from the bottom layer, $h(x)$.  
Its scaling properties are studied by calculating the 
root-mean-square surface width and the generalized increment correlation 
functions $C_q(x)= \langle |h(x_0+x)-h(x_0)|^q \rangle$. 
The surface is found to have a nontrivial scaling behavior 
on small length scales, with a crossover to scale-independent 
behavior on large scales. 
As the vacancy concentration approaches the site-percolation limit, both the 
crossover length and the saturation value of the surface width diverge in a 
manner that appears to be proportional to the bulk connectivity length. 
This suggests that a percolation transition in the bulk 
also drives a similar divergence observed in surfactant 
templated polyacrylamide gels at high surfactant concentrations. 
\end{abstract}

\maketitle

\section{Introduction}
\label{sec:intro}

Crosslinked polymer hydrogels are completely interconnected
polymer networks that combine high water content with high
porosity, forming macroscopic molecules that have applications in
many fields. Their wide range of pore sizes makes such gels ideal
for separation of biological macromolecules by electrophoresis or
chromatography \cite{Gen1}. The crosslinking process induces a
reorganization of the polymer structure, resulting in  
inhomogeneities in the spatial density 
\cite{COHE92,Osh-Bengui96,inhom2,inhom1,inhom3}. 
These inhomogeneities
affect the surface configurations of the gels, but only in recent
years have advances in atomic force microscopy (AFM) made possible
the imaging of soft material surfaces in an aqueous environment
\cite{Suzuki95}. The effects of the crosslinking density,
temperature, pressure, and sample thickness on the surface
topography have been studied by Suzuki {\it {et al}}.
\cite{Suzuki95,Suzuki97}. Their results indicate that the structural
features of the surface on both the micrometer and nanometer
scales depend on these factors. It has been suggested
\cite{Suzuki95} that control of the characteristic length scale of
the gel surface using external stimuli may have applications in a
variety of fields, such as regulation of adsorption and release of
specific molecules by the intermolecular forces between the
surface and the molecule. Thus, gel surfaces provide a unique
opportunity to explore the interplay between phenomena on the
macroscopic and nanoscopic scales.

Templated polyacrylamide gels are formed by polymerizing
acrylamide with a crosslinker in the presence of a surfactant.
The surfactant molecules form monodisperse micelles 
of a size roughly comparable with the
crosslinker separation, and the presence of these micelles alters the
gel pore structure, enhancing the gel's separation properties
\cite{rill96}. In a recent work the surface morphology of
templated polyacrylamide gels was extensively studied by AFM 
and scaling analysis of the resulting images \cite{Mukundan01}. This study
indicates that the gel surfaces are self-affine on short length
scales, with roughness (Hurst) exponents on the order of 0.8--1.
In the absence of surfactant a cross-over length, above which the
surface is no longer self-affine, was estimated to be on the order
of 300~nm, and the saturation value of the interface
width was on the order of 1~nm. Both values increased dramatically
with the introduction of surfactant.

Detailed kinetic lattice models of the polymerization of crosslinked
polymer gels have previously been constructed \cite{Pandey}.
However, they emphasize the kinetics of polymerization and are
extremely computationally intensive. Inspired by the experiments reported in 
Ref.~\cite{Mukundan01}, the aim of the present work
is rather to construct and study a simple continuum model that can
reproduce some of the observed scaling characteristics of the 
templated gel surfaces at a more modest (but still substantial) computational
cost by concentrating on the elastic structure of the gel. Some
preliminary results were presented in Ref.~\cite{UGA01}.

The remainder of this paper is organized as follows. In
Sec.~\ref{sec:model} we introduce the model and detail some aspects of the
numerical calculations. In Sec.~\ref{sec:scal} we recall some
scaling concepts associated with the analysis  of surfaces. In
Sec.~\ref{sec:results} we show the results of our calculations,
in particular the scaling properties of the surface
width and increment correlation function 
along with their dependence on the size of the sample. 
We also comment on the relevance of our results 
for the interpretation of recent AFM experiments. Finally, in
Sec.~\ref{sec:conc} we present our conclusions.

\section{Model}
\label{sec:model}

The model consists of a two-dimensional network of nodes
interconnected by  massless springs. The nodes represent
crosslinker molecules, and the connecting springs represent
polymer chains. The network topology consists of a 
triangular lattice of nodes, each of which is
connected by harmonic springs to its six nearest neighbors (except
at the top and bottom surfaces, where each node has only
four connections). The model has no excluded-volume interactions. 
A triangular lattice was chosen to
ensure geometrical stability in two dimensions without introducing
bond-angle constraints. The corresponding unrealistically high crosslinker 
functionality should not significantly influence our results. 
The network has periodic boundary
conditions in the horizontal ($x$) direction, free boundary conditions
at the top ($y>0$) layer, and the nodes in the bottom layer are fixed at $y=0$,
corresponding to bonding to a rigid substrate. There are $L_x$
nodes in the horizontal direction and $L_y$ nodes in the vertical
direction.

The total energy of the network is $E=\sum_i (1/2) k_i (
l_i-l_{0i} )^2$, where $k_i$, $l_i$, and $l_{0i}$, are the spring
constant, the actual length, and the equilibrium length under zero external 
force of the $i$th spring, respectively. (All quantities in this paper are
given in dimensionless units.) The equilibrium length of each
spring is independent of the other springs and is randomly chosen
with probability density function (pdf) 
\begin{equation}
P(l_{0i})=2 \gamma l_{0i} \exp \left (- \gamma l_{0i}^2\right),
\label{eq:lprob}
\end{equation}
where $\gamma$ is proportional to the inverse of the average
number of monomers between crosslinkers. This pdf 
corresponds to the case that the equilibrium distance between crosslinkers is 
proportional to the square root of an exponentially distributed
number of monomers. It is consistent with the
picture that crosslinkers are distributed randomly along the polymer
chains, and that a spring of equilibrium length $l_{0i}$ corresponds to a
polymer of the same average end-to-end distance in the random-coil
collapsed phase \cite{RICH71,Text}. The average equilibrium length of a
spring is $\langle l_{0i} \rangle = \sqrt{{\pi}/{\gamma}}/2$. In
agreement with a mean-field spin-chain approximation for the
elastic properties of a collapsed polymer chain \cite{Text}, we
require that the elastic constant of the $i$th spring should be inversely
proportional to its equilibrium length, $k_i = l_{0i}^{-1}$ in our 
dimensionless units.

With the above distribution of equilibrium lengths,
the pdf for the force $F$ exerted by a spring of length $l$ is
\begin{equation}
P(F|l) = \frac{2 \gamma l^2}{(1-F)^3}
\exp \left[ - \gamma \left( \frac{l}{1-F}\right)^2 \right]
\label{eq:fprob}
\end{equation}
for $F \le 1$, while it vanishes for $F>1$. 
The average force exerted by a spring of length $l$ is then
\begin{equation}
\langle F|l \rangle =  1 - l \sqrt{\gamma \pi} \;.
\label{eq:avgf}
\end{equation}

Initially, springs with equilibrium lengths drawn independently
according to $P(l_0)$ are placed on the bonds of a regular triangular
lattice of unit lattice constant, such that all of  them are
stretched or compressed to a length of one. In order that
this initial configuration should not be globally stressed, we impose the
condition that the average force exerted by a spring of unit
length should be zero. By Eq.~(\ref{eq:avgf}), this
is satisfied for $\gamma=1/\pi$. However, the initial
configuration is {\it locally} stressed, since $l_i=1$ for all $i$
while $\langle l_{0i} \rangle = \pi/2 \approx 1.57$. 
The system is frustrated in the sense that there is no configuration that 
can simultaneously minimize the energies of all the springs. As a result 
there is a large number of configurations corresponding to local energy 
minima with similar energies. 

The network was relaxed using the limited-memory 
Broyden-Fletcher-Goldfarb-Shanno (L-BFGS) quasi-Newton minimization
algorithm \cite{subr,LBFGS} until a locally stable configuration was
reached, corresponding to a local minimum in the global energy
landscape. By minimizing several statistically equivalent 
realizations of the random spring configuration,  
we checked that the local minima reached by this
procedure are representative of typical relaxed configurations.
Although the total energies of different relaxed configurations
were not completely identical, all the structural results were the
same to within our numerical accuracy. 

The L-BFGS subroutine requires the calculation of the total energy and its
gradients, i.e., the local forces in the vertical and horizontal
directions for each node. By monitoring the total energy and the
absolute value of the force we found that the subroutine must be
called about $3 \times 10^4$ times to ensure that the system
reaches a local minimum. 
Since each relaxation step requires a new
evaluation of the total energy and the local forces, the L-BFGS
algorithm is quite computationally intensive, taking about 1-2 days
on a current 600~MHz dual-processor Pentium III PC with 512~MB of shared memory 
to relax a typical system with
about $10^6$ degrees of freedom. However, 
it was far more efficient for our application than a na{\"\i}ve
steepest-descent method, requiring about an order of magnitude
less computer time while leading to essentially equivalent final
energies and structures.
After relaxation, we examined the bulk and surface properties of the 
network. Images of a typical relaxed network are shown in
Figs.~\ref{fig:zooms} and~\ref{fig:bulk}. Inhomogeneities in the
spatial crosslinker density are evident in these figures. 

\section{Scaling properties of surfaces}
\label{sec:scal} 
The scaling properties associated with the height
fluctuations of rough surfaces have been the object of a great
deal of interest. Vapor deposition, epitaxial growth, polymer
growth, and many other phenomena exhibit scale invariance
\cite{Vicsek,yang93,barabasi95,meakin98}.  In a statistical sense, 
the height of a single-valued
self-affine random surface, $h(x)$, satisfies the scaling relation
\begin{equation}
h(x)=\lambda^{-H}h(\lambda x)\;,
\label{eq:hxscal}
\end{equation}
where $\lambda$ is a dimensionless scaling parameter and $H$ is the so-called
Hurst exponent, which in this case is identical to the roughness exponent
\cite{meakin98}. Consequently, the increment correlation function
$C(x)$ of a translationally invariant, self-affine surface scales as
\begin{equation}
C(x)=\langle [h(x_0+x)-h(x_0)]^2 \rangle \sim x^{2H}\;,
\label{eq:Cx}
\end{equation}
where the average is taken over $x_0$. 
Many surfaces present a mixed behavior characterized by
\begin{equation}
C(x)  \sim
\left\{ \begin{array}{ll}
x^{2H} & \mbox{for $x << l_{\times}$} \\
 \mathrm{constant} & \mbox{for $x >> l_{\times}$}
\end{array}
\right.
\;,
\label{eq:cross}
\end{equation}
where $l_{\times}$ is a cross-over length scale separating the two
behaviors. However, there exist surfaces for which the scaling
cannot be described by a single exponent; instead an infinite
hierarchy of characteristic exponents has to be introduced. 
These surfaces are called multi-affine, and they
are occasionally observed in nature \cite{barabasi91,barabasi92,barabasi92B}.
Molecular-beam epitaxy (MBE) models \cite{Das}, polished metal surfaces 
\cite{Dharma}, geomorphological features \cite{KIMKONG00}, 
and time series of financial data \cite{IVAN99,KATS00} provide some examples of 
multi-affine scaling. The characteristic exponents of a
multi-affine function can be calculated from the $q$th-order
increment correlation function, defined as
\begin{equation}
C_q(x)= \langle |h(x_0+x)-h(x_0)|^q \rangle \sim x^{qH_q}
\label{eq:Cq}
\end{equation}
with the generalized Hurst exponent $H_q$ depending continuously
on $q$, at least in some region of $q$ values. Thus $C_2(x)$ is
identical to the usual increment correlation function $C(x)$
defined in Eq.~(\ref{eq:Cx}). Like self-affine surfaces, 
multi-affine surfaces can show a cross-over to scale-independent behavior, 
analogous to Eq.~(\ref{eq:cross}). 

In the next section we present our results
on the structure of the bulk and the scaling behavior of the
surface, calculated after the spring network has been relaxed to a
representative local energy minimum.

\section{Numerical Results}
\label{sec:results}

We analyzed spring networks of sizes $L_x \times L_y$. In order 
to simulate experimental measurements of the scaling behavior of surfaces 
of gels in which pores have been introduced by polymerizing
acrylamide plus a crosslinker templated with surfactants that
are later removed \cite{rill96,Mukundan01}, we created
voids in the networks. The voids were introduced by selecting a
proportion of the nodes at random and removing these nodes 
and the springs associated with them. Next, the 
Hoshen-Kopelman algorithm \cite{Perco,STAU91} 
was applied to identify and discard 
clusters that were not connected to the fixed substrate. This last step is 
particularly important near the site-percolation limit, which for the 
triangular lattice is at a vacancy concentration of 50\% \cite{STAU91,SMIR01}. 
Note that by ``vacancy concentration" we always mean the initial concentration 
of vacant nodes before removal of the disconnected clusters. The total volume 
fraction of the void structure resulting from the 
cluster removal is of course larger. 

We define the one-dimensional surface as the set of surface heights, 
$h(x_j)$, at equally spaced discrete points, $x_j \in [0,L_x)$,
considering the polymer chains as straight line segments between
nearest-neighbor connected nodes. To calculate the surface height at
$x_j$, we first identify the set of all springs which intersect a
vertical line at $x_j$. From this set of springs, we define
\begin{equation}
h(x_j)=\max{[y_i(x_j)]},
\label{eq:surfdef}
\end{equation}
where $y_i(x_j)$ is the vertical  location of the intersection of
the $i$th spring  at $x_j$. We use the sign convention that $y=0$
for the fixed bottom nodes and the free surface is located at 
positive values of $y$. In Fig.~\ref{fig:zooms} we show details of
the bulk and free surface of systems with vacancy concentrations of
$0$\%, $40$\%, and $49.5$\%, while in Fig.~\ref{fig:bulk} we show a magnified 
view of a section of the bulk of a vacancy-free system. 

To obtain an estimate of the finite-size effects and select an
appropriate system size for the production runs, we first considered 
vacancy-free spring networks of sizes $L_x \times L_y $,  with
$L_x=512$, 1024, and $2048$, and $L_y=256$, 384, 512, 640, and
$748$. We calculated the root-mean-square width of the surface
(also known as the rms roughness) on a length scale $L$, $w_L$, 
by recursively subdividing the
surface into $n$ nonoverlapping segments of length 
$L=L_x/n$, with $n$ an integer between 2 and $L_{x}/2$. The mean-square 
width of the $\it k$th segment is 
\begin{equation}
w_{Lk}^2 = \langle y^2 \rangle_{Lk} - \langle y \rangle_{Lk}^2 \;,
\label{eq:rmsdef}
\end {equation}
where $\langle \cdot \rangle_{Lk}$ denotes averaging within the
segment. The rms width is defined as the square root of the average of 
$w_{Lk}^2$ over the $n$ segments, 
$w_L = \sqrt{{n^{-1}\sum_{k=1}^{n}{w_{Lk}^2}}}$.
The statistical errors in $w_L$ were estimated in the usual way
from the standard deviation of $w_{Lk}^2$. 

At this point we 
emphasize that the order in which the average and the square root
are taken in the definition of $w_L$ 
is important. For a simple self-affine surface it is
irrelevant whether the square root is taken before or after the
average; in either case one obtains the same exponent. However,
for a multi-affine surface the results are quite different. In
this case the exponent obtained from $w_L$ is equal to that
calculated from the increment correlation function $C(L)$ only
when the average is calculated {\em before\/} taking the square root, 
as it is done here. The results for $w_L$, averaged over four independent 
configurations for each system size, are shown in Fig.~\ref{fig:size}. 

The data shown in Fig.~\ref{fig:size}
indicate that, for the values of $L_x$ and $L_y$ considered, $w_L$ is
independent of the system size. However, when comparing the
average height for the $j$th layer in the
relaxed system, $\langle y_j \rangle$, 
with its original value in the regular triangular
lattice, we found a slight contraction limited to approximately
the top 250 layers. To minimize finite-size effects due to interference of
the finite thickness $L_y$ with this skin effect, while
maintaining $L_x$ as large as possible so as to reach large horizontal
length scales and reduce statistical uncertainties in the
estimates of $w_L$, we chose a system size of 
$L_x \times L_y = 1024 \times 768$ for the production runs.
Both the bulk and surface structures were analyzed as discussed
in the following two subsections.

\subsection{Bulk structure}
\label{bulk}

In this subsection, we discuss the bulk properties of the spring
network far from both the fixed and free surfaces. We only present 
results for a gel without vacancies. For gels with high
vacancy concentrations, it is difficult to define a bulk region,
since the width of the surface region increases with increasing
vacancy concentration. 
To study the bulk structure in this case, simulations with periodic boundary 
conditions would be preferable. 

\begin{figure}
{\epsfxsize=2.5in \epsfbox{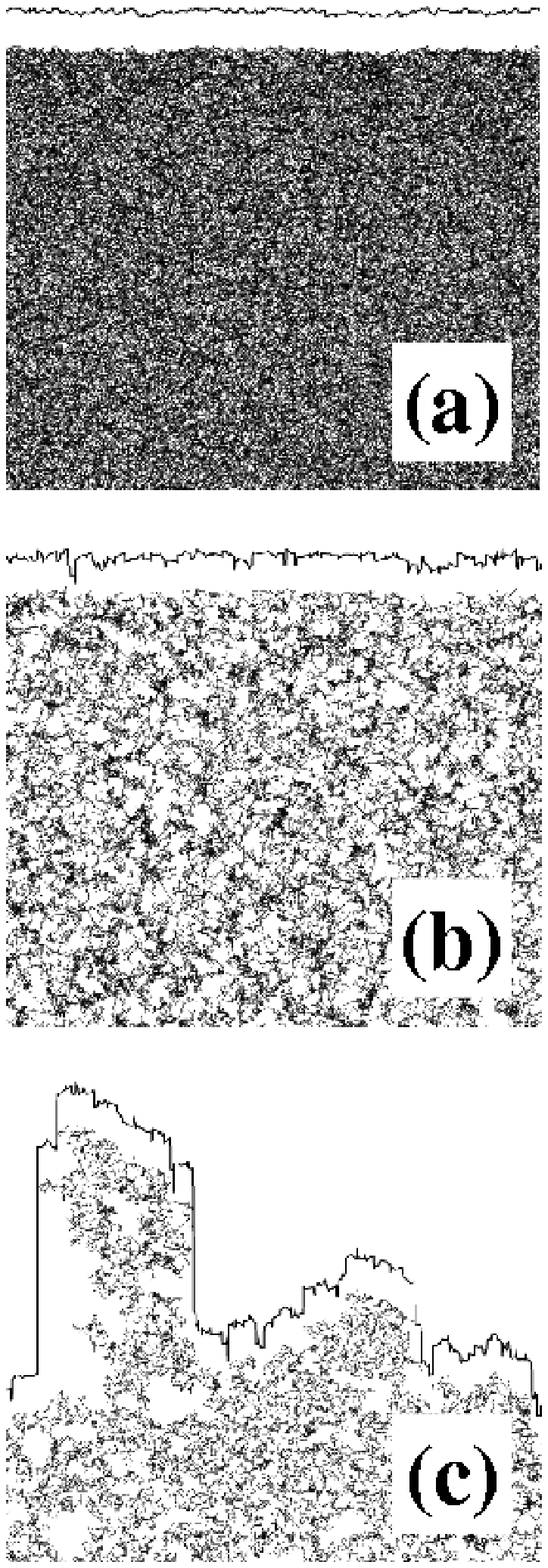}} 
\caption[]{
Details of the surface and bulk structures of relaxed $1024 \times 768$
systems with vacancy concentrations of (a) 0\%, (b) 40\%, and (c)
49.5\%. The thin line segments represent polymer chains. The bold
lines (vertically displaced for easier viewing) represent the free
surface.}
\label{fig:zooms}
\end{figure}

\begin{figure}
{\epsfxsize=3.4in \epsfbox{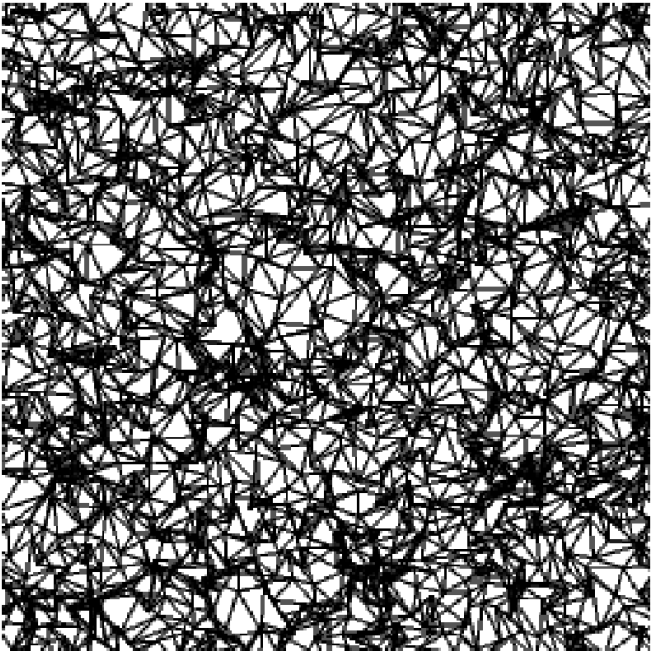}} 
\caption[]{Magnified detail
of the relaxed bulk structure for a 
$1024 \times 768$ system without vacancies.}
\label{fig:bulk}
\end{figure}

\begin{figure}
{\epsfxsize=3.4in \epsfbox{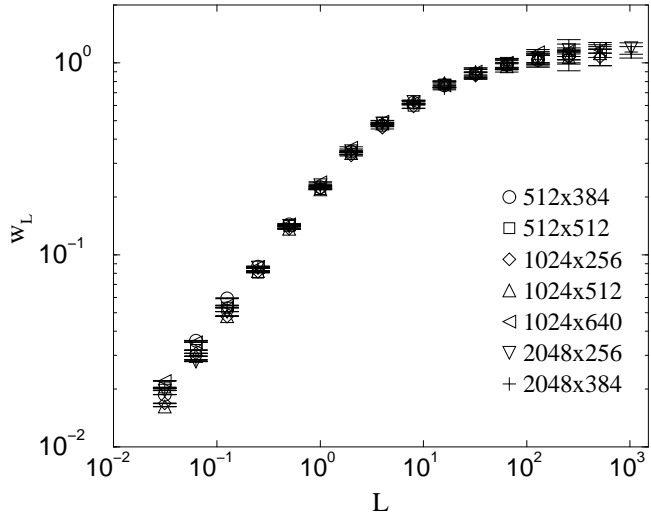}} 
\caption[]{The rms surface
width $w_L$ for vacancy-free spring networks, 
shown on a log-log scale as a function of the length
scale $L$ for a number of system sizes. The two scaling
regimes defined in Eq.~(\ref{eq:cross}) are clearly seen. 
Averaged over four independent configurations for each system size.}
\label{fig:size}
\end{figure}

The presence of inhomogeneities in the spatial  crosslinker
distribution, where the crosslinkers and the polymer chains (shown as line
segments) have greater densities in some regions (clumps), can be
seen in Fig.~\ref{fig:zooms} and Fig.~\ref{fig:bulk}. 
The effect is more pronounced for 
higher vacancy concentrations. This is a desirable feature of
the model since static inhomogeneities are characteristic of crosslinked
polyacrylamide gels and have been related to a freezing-in of the
topological structure during polymerization 
\cite{COHE92,Osh-Bengui96,inhom2,inhom1,inhom3}. For 
gels without vacancies, Fig.~\ref{fig:zooms}(a),
the clumping is not as obvious to the eye. Thus, we show in
Fig.~\ref{fig:bulk} a magnified image of a small region of the
bulk of a vacancy-free system, in which the clumping can be clearly seen. 
To quantify the clumping, we characterize the bulk scaling
properties of the gel by a method analogous to the width-scaling
method of Eq.~(\ref{eq:rmsdef}).

\begin{figure}
{\epsfxsize=3.4in \epsfbox{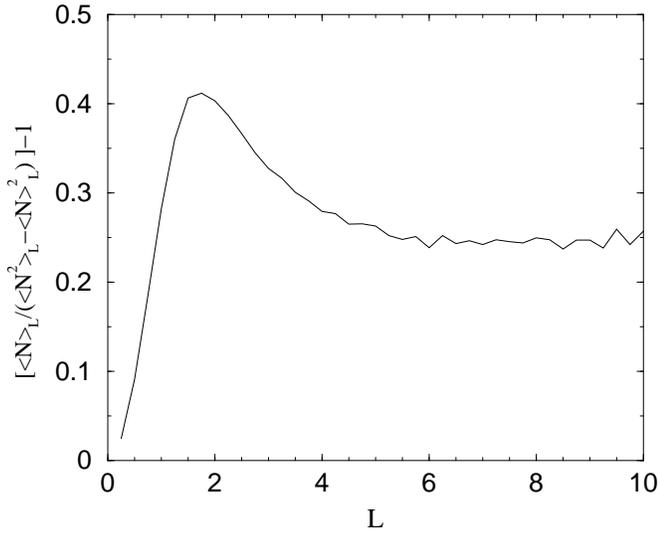}} 
\caption[]{The bulk scale
parameter vs the length scale $L$. The data are for a system of
$1024 \times 768$ nodes with 0\% vacancies, averaged over eight
independent configurations. The function values near zero
indicate randomly placed nodes at small length scales, and the
maximum at $L \approx 1.57$, which is the average equilibrium spring length,
indicates that nodes are most regularly placed on this scale.
}
\label{fig:avgbulk}
\end{figure}

We define $\rho(x,y)=\sum_i \delta(x-x_i,y-y_i)$ to be the density
of nodes at a particular point $(x,y)$ of the bulk. Here
$\delta(\cdot)$ is the two-dimensional Dirac delta function, and
the sum runs over all nodes $(x_i,y_i)$ in the system. We calculate the mean
and the variance of the number of nodes inside a box of side $L$,
$\langle N \rangle_L$ and $\langle N^2\rangle_L -\langle N
\rangle_L^2$, respectively, by averaging $\rho(x,y)$ over
non-overlapping boxes. When there are no correlations within the
gel, corresponding to a purely random distribution of node
positions, the number of nodes in a box obeys a Poisson
distribution. This yields $\langle N \rangle_L/(\langle
N^2\rangle_L -\langle N \rangle_L^2)=1$. However, when the
nodes have a more regular distribution, the denominator is reduced
and the ratio increases. In Fig.~\ref{fig:avgbulk} 
$\langle N \rangle_L/(\langle N^2\rangle_L -\langle N
\rangle_L^2)-1$, averaged over eight independent realizations of the system, 
is shown vs $L$ from $L=0.25$ to $10$. For small 
$L$, the function tends to zero, indicating that nodes which are
very close together are essentially randomly placed. The function 
goes through a maximum near the average spring length ($L \approx
1.57$), corresponding to a more regular distribution of the nodes
on this scale.

\subsection{Surface scaling behavior}
\label{Surface} 

\begin{figure}
{\epsfxsize=3.3in \epsfbox{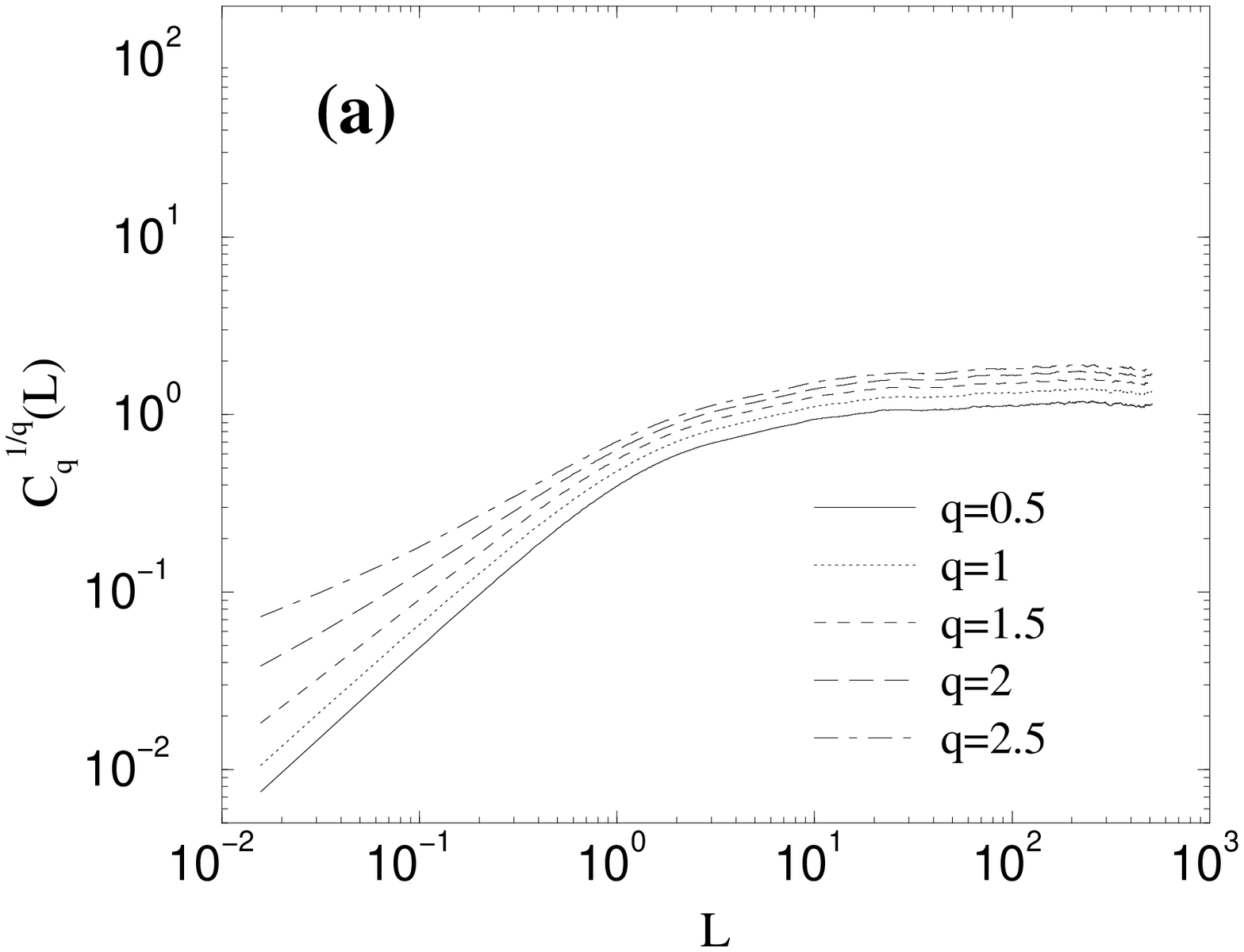}} {\epsfxsize=3.3in
\epsfbox{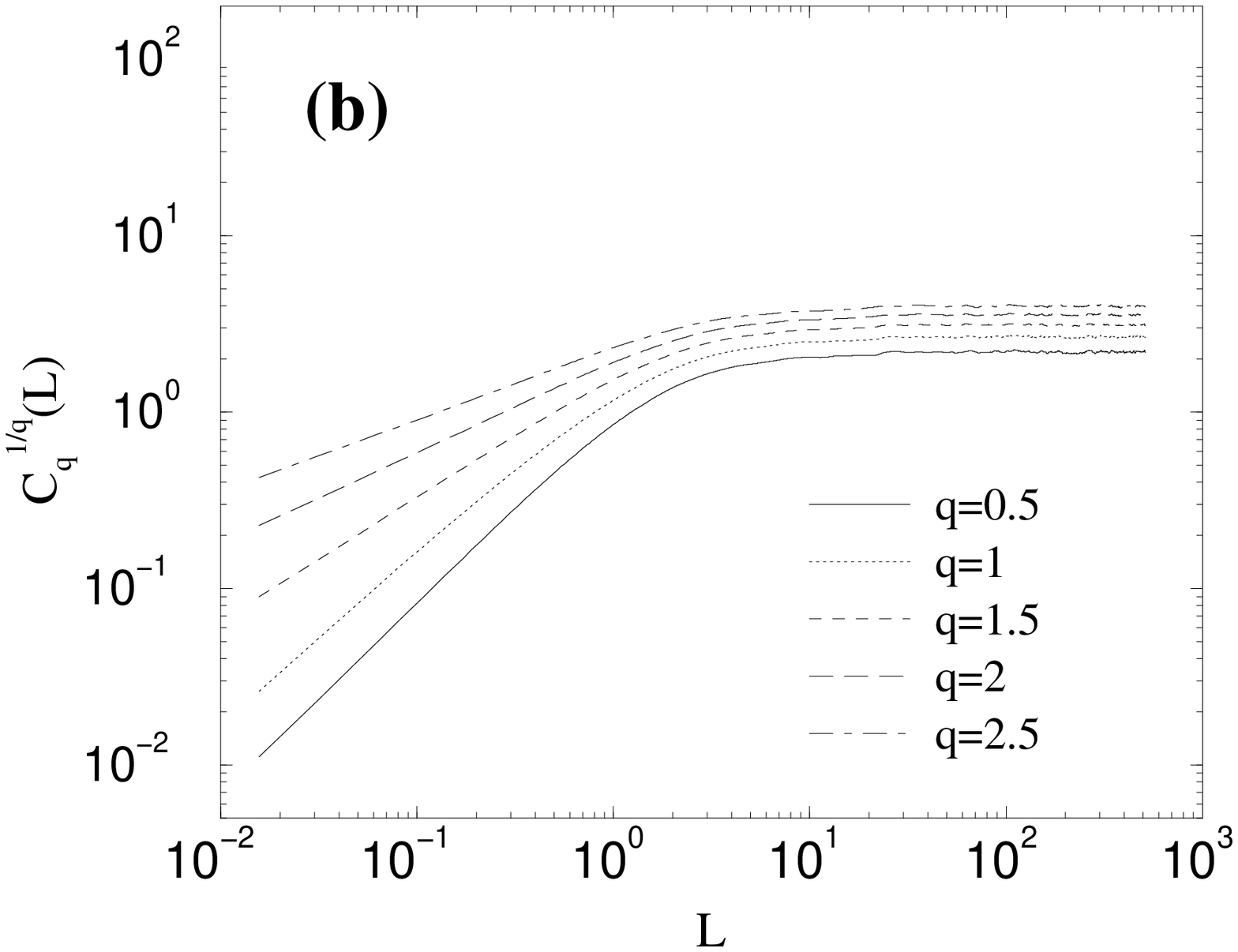}} {\epsfxsize=3.3in \epsfbox{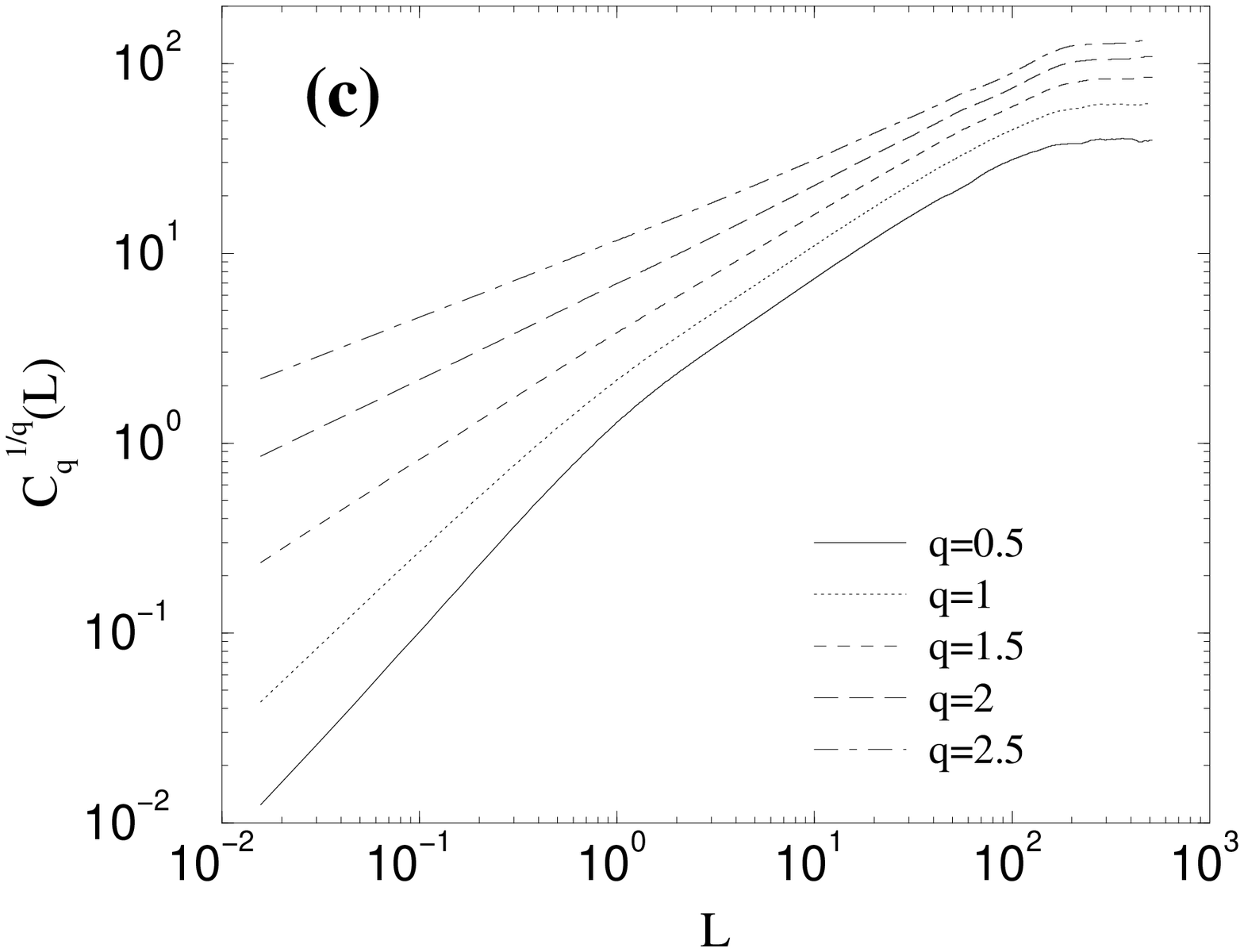}}
\caption[]{The $q$th root of the generalized increment correlation
function $C_q(L)$ for systems with vacancy concentrations 
(a) 0\% vacancies, averaged over ten independent realizations. 
(b) 40\% vacancies, averaged over ten independent realizations. 
(c) 49.5\% vacancies, averaged over thirteen independent realizations.
}
\label{fig:q0}
\end{figure}

We calculated the $q$th root of the $q$th-order
increment correlation function, $C^{1/q}_q(L)$, for $q=0.5$, 1,
1.5, 2, and $2.5$. The results for systems
with $0$\%, $40$\%, and $49.5$\% vacancies are shown in
Fig.~\ref{fig:q0}(a), (b), and (c), respectively. For each of the two lower 
vacancy concentrations the data were averaged over ten 
independent realizations of the system, 
while thirteen realizations were used for the highest concentration. 
For all three vacancy concentrations, the surface is seen to be multi-affine 
with $q$-dependent slopes for $L$ 
below a crossover length, $l_{\times}$. [For the highest vacancy concentration 
(Fig.~\ref{fig:q0}(c)), two scaling regimes with different, nonzero $H_q$ 
are seen for $q \le 1$.] For $L > l_{\times}$, $C^{1/q}_q$ reaches
a saturation value $C^{1/q}_{\rm sat}$. Both of these length scales, 
$l_{\times}$ and $C^{1/q}_{\rm sat}$, depend on 
the vacancy concentration and $q$ in ways that are discussed below. 

\begin{figure}
{\epsfxsize=3.2in \epsfbox{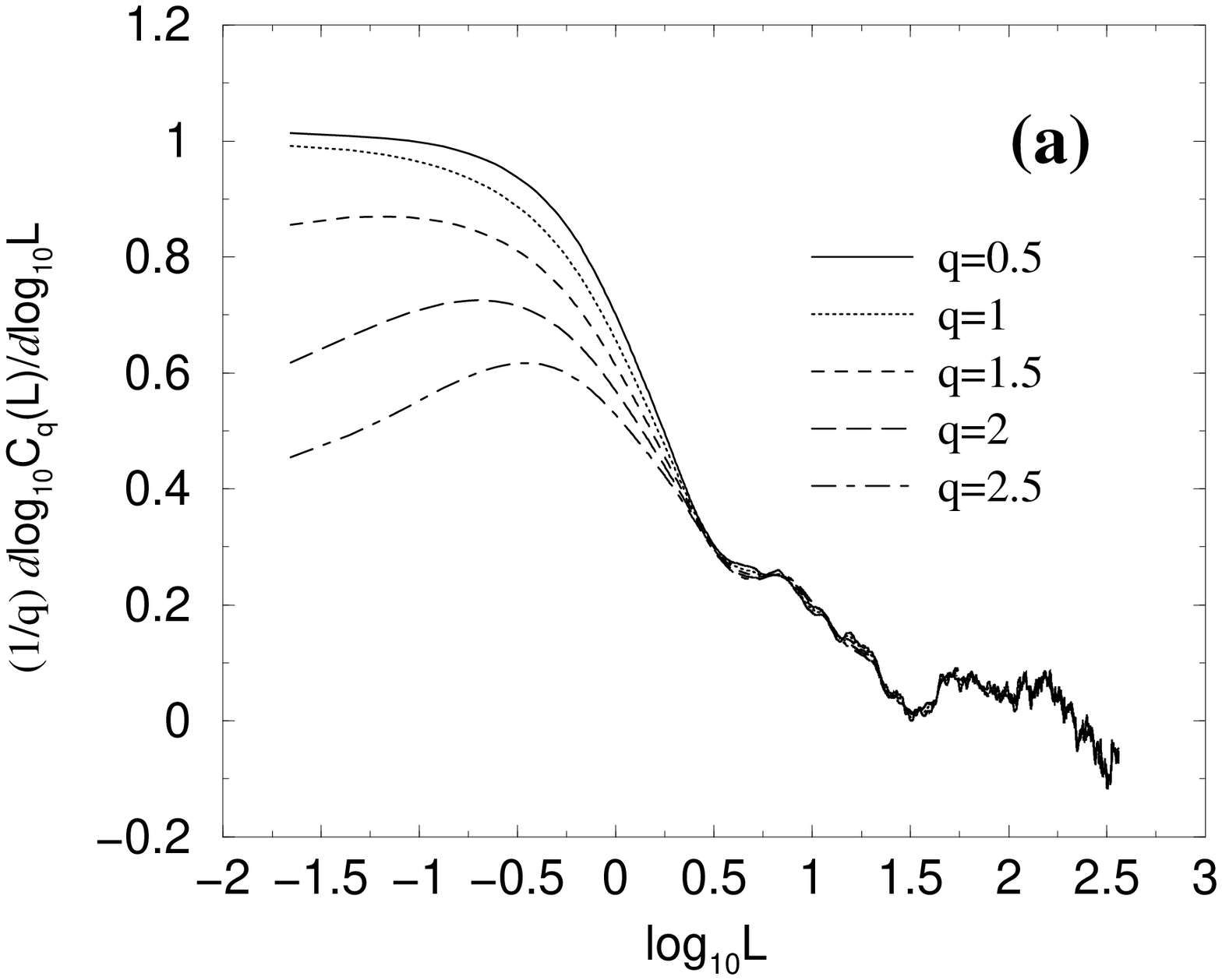}} 
{\epsfxsize=3.2in \epsfbox{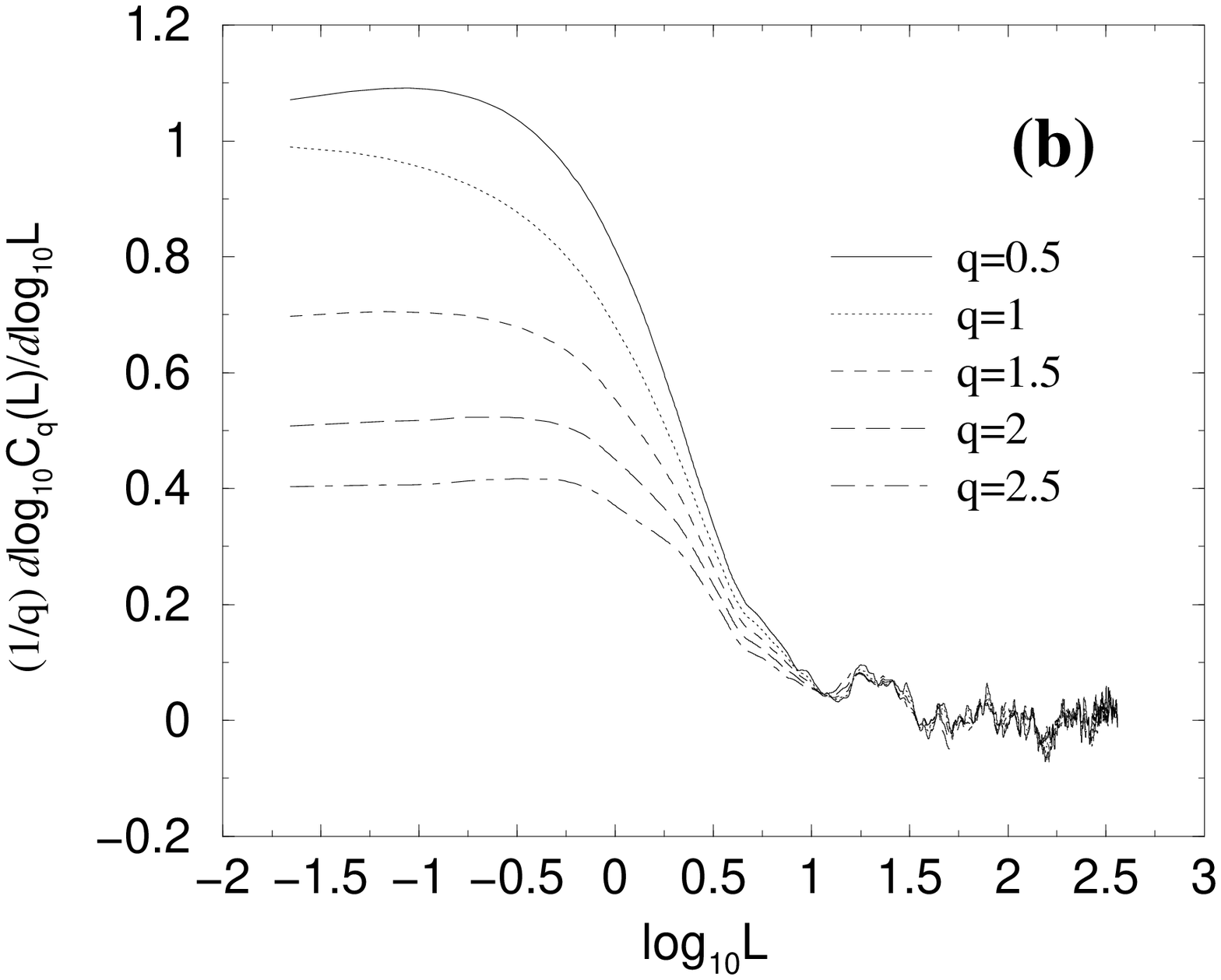}} 
{\epsfxsize=3.2in \epsfbox{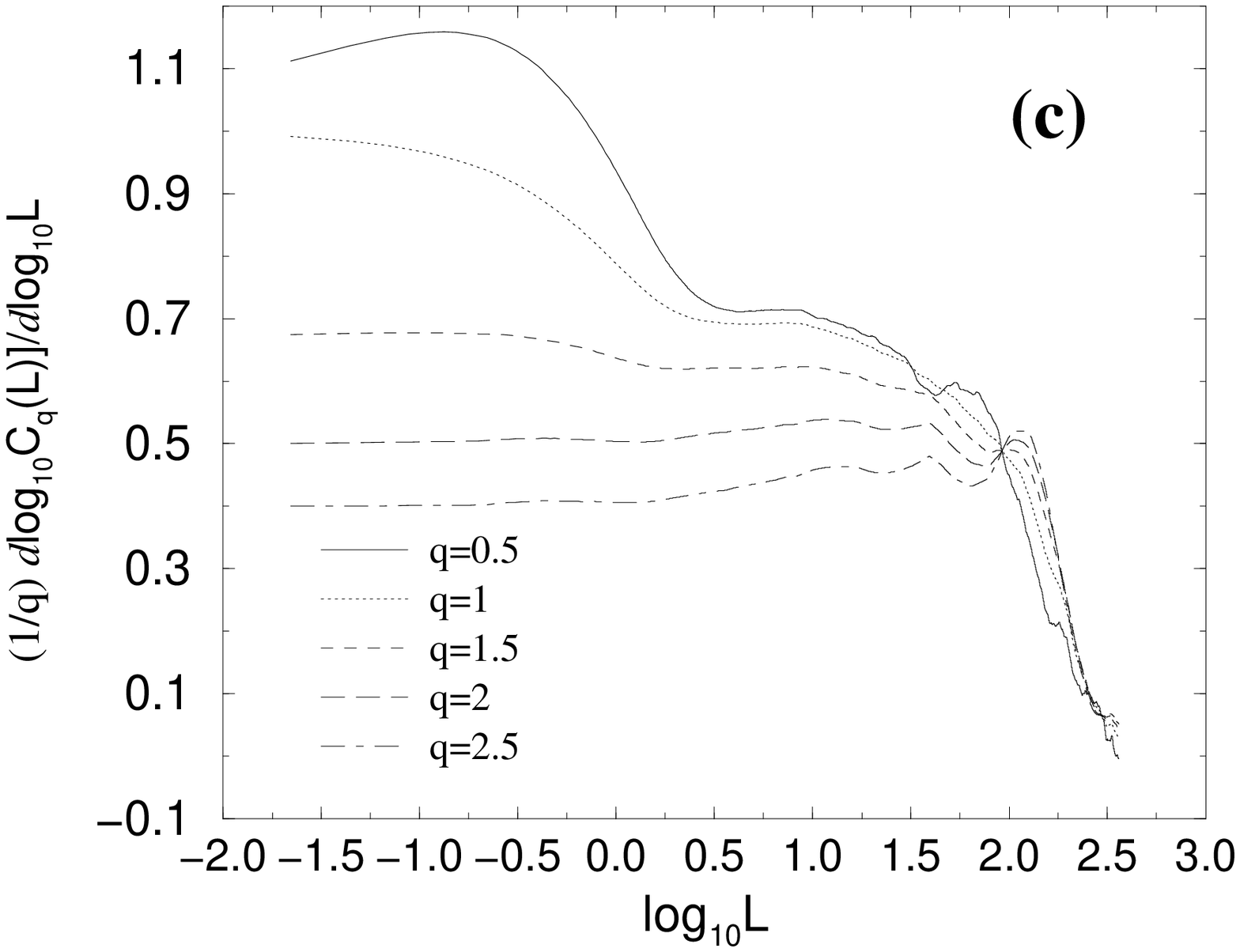}}
\caption[]{The effective generalized Hurst exponent $H_q$
for systems with different vacancy concentrations, obtained as
logarithmic two-point derivatives of $C_q^{1/q}(L)$ with respect
to $L$. The data for $C_q^{1/q}(L)$ were averaged as in 
Fig.~\protect\ref{fig:q0}. 
For small $L$ and $q$, the derivative approaches unity,
indicating $H_q \approx 1.0$. For large $L$, $H_q$ 
approaches zero, signaling the scale-independent 
saturation regime. 
(a) 0\% vacancies.
(b) 40\% vacancies.
(c) 49.5\% vacancies.
}
\label{fig:dq0}
\end{figure}

\begin{figure}
{\epsfxsize=3.4in \epsfbox{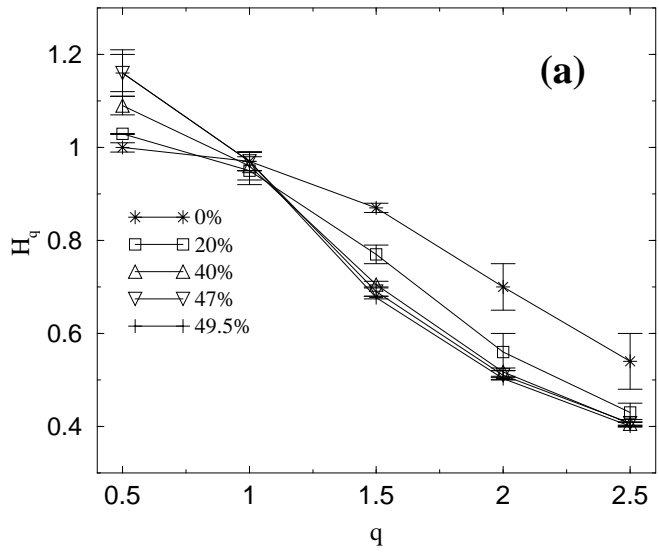}} 
{\epsfxsize=3.4in \epsfbox{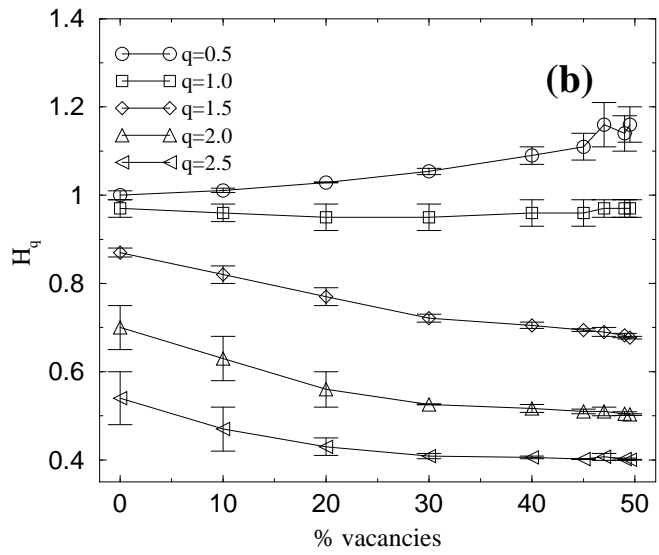}} 
\caption{The generalized Hurst
exponent $H_q$ for systems with different vacancy concentrations, from 
Table~\ref{tab:exp}. 
(a) Shown vs $q$ for different vacancy concentrations. 
(b) Shown vs vacancy concentration for different values of $q$. 
In both parts the lines are merely guides to the eye. 
}
\label{fig:hq}
\end{figure}

By numerically calculating the logarithmic derivatives of the
correlation functions with respect to $L$ we determine the scaling exponents, 
\begin{equation}
H_q
= 
\frac{1}{q}\frac {d \log \, C_q(L)}{d \log L} 
\;.
\label{eq:Hq}
\end{equation}
Effective values for $H_q$ for different $L$ and 
$q$ were estimated as two-point derivatives over intervals
corresponding to a doubling of $L$. The results are shown in
Fig.~\ref{fig:dq0}(a), (b), and (c) for systems with $0$\%,
$40$\%, and $49.5$\% vacancies, respectively. The figures clearly
indicate that on length scales below $l_{\times}$, the
surfaces are multi-affine with $q$-dependent scaling exponents $H_q$. 
In this scaling regime the effective exponents are approximately independent 
of $L$. Scaling exponents $H_q$ for different values of the vacancy
concentration are presented in Table~\ref{tab:exp}. The exponents
shown in this table were obtained from Eq.~(\ref{eq:Hq}) at fixed $L=0.088$, 
in the small-$L$ scaling regime. In Fig.~\ref{fig:hq} we show how these 
values of $H_q$
change with $q$ and the vacancy concentration. For $q > 1$, $H_q$ decreases 
with increasing vacancy concentration, while it increases for $q < 1$. For 
$q=1$, $H_q$ is slightly below unity and 
independent of the vacancy concentration to within our numerical accuracy. 

\begin{figure}
{\epsfxsize=3.2in \epsfbox{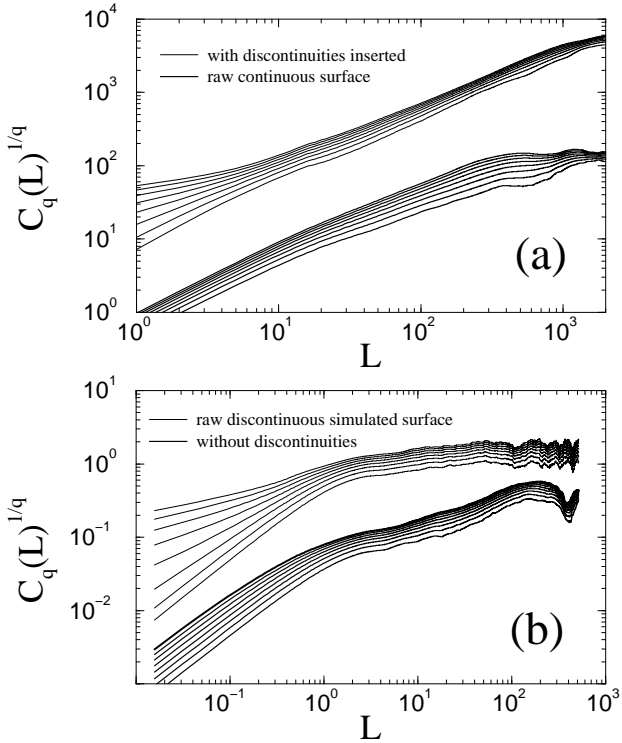}} 
\caption[]{
Effects of discontinuities on the generalized correlation functions
$C_q(L)^{1/q}$ and their logarithmic slopes, the generalized Hurst
exponents $H_q$, for self-affine and multi-affine surfaces. In all sets of
curves shown, $q$ goes from 0.5 to 4.0 in
increments of 0.5, from below to above in the figures. 
For clarity of view, the data corresponding to continuous surfaces 
(the thick curves) have been divided by ten. 
(a) 
The thick curves correspond to continuous, 
approximately self-affine surfaces, as
indicated by the $q$-independent $H_q$. 
The thin curves represent discontinuous surfaces obtained by inserting a
finite density of vertical discontinuities in the original surfaces. 
For the latter surfaces, $H_q$ depends on $q$, indicating 
multi-affinity on length scales less than ten.  
Each curve is an average over data for 1000 surfaces. 
(b)
The thin curves correspond to a multi-affine 
simulated gel surface with 0\% 
vacancies, corresponding to those for which $C_q(L)^{1/q}$ 
are shown in Fig.~\ref{fig:q0}(a). 
The thick curves correspond to a continuous surface generated from the
simulated gel surface by removing all vertical discontinuities. 
In this case, $H_q$ is independent of $q$, indicating simple self-affinity. 
}
\label{fig:discont}
\end{figure}

Multi-affinity is relatively rare in natural surfaces, and, in fact, 
it is not observed in the experimental system that inspired the present 
study \cite{Mukundan01}. It is therefore reasonable to inquire about the source 
of the multi-affinity observed in our model system. We find that it is 
caused by the finite density of discontinuities, which result from overhangs in 
the simulated surfaces (see Fig.~\ref{fig:zooms}). This effect is illustrated in 
two ways in Fig.~\ref{fig:discont}. In Fig.~\ref{fig:discont}(a) we show 
on a log-log scale $C_q(L)^{1/q}$ for different values of $q$ for 
surfaces generated by fitting linear line segments of unit length to 
self-affine surfaces with a Hurst exponent of 0.75, 
which were generated by  the method of successive additions \cite{VOSS}.
These surfaces are seen to be self-affine with a Hurst exponent which
crosses over from near unity to approximately 0.75 at about $L=1$. 
In the same figure we also show (by thin curves) the corresponding
generalized increment 
correlation functions for surfaces generated from the above
ones by the insertion of a finite density of vertical discontinuities. In
contrast to the result for the original surfaces, the
slopes of the latter curves depend on $q$, indicating that the
discontinuous surfaces are multi-affine. 
Conversely, in  Fig.~\ref{fig:discont}(b) we illustrate the effect of
removing the discontinuities from a surface generated by our model
with 0\% vacancies. While the original surface is 
multi-affine (signified by the $q$-dependent slopes), the surface from
which the discontinuities are removed is self-affine with a Hurst
exponent of approximately 0.98, corresponding to the straight line segments 
used to represent the polymer chains. 
These numerical results clearly show that the multi-affinity in our model
surfaces is due to the vertical discontinuities caused by the overhangs. 
Possible methods to improve the agreement of the scaling behavior 
of the model surfaces with that of the experimental system by removing the 
discontinuities are discussed in Sec.~\ref{sec:conc}. 

\begin{figure}
{\epsfxsize=3.4in \epsfbox{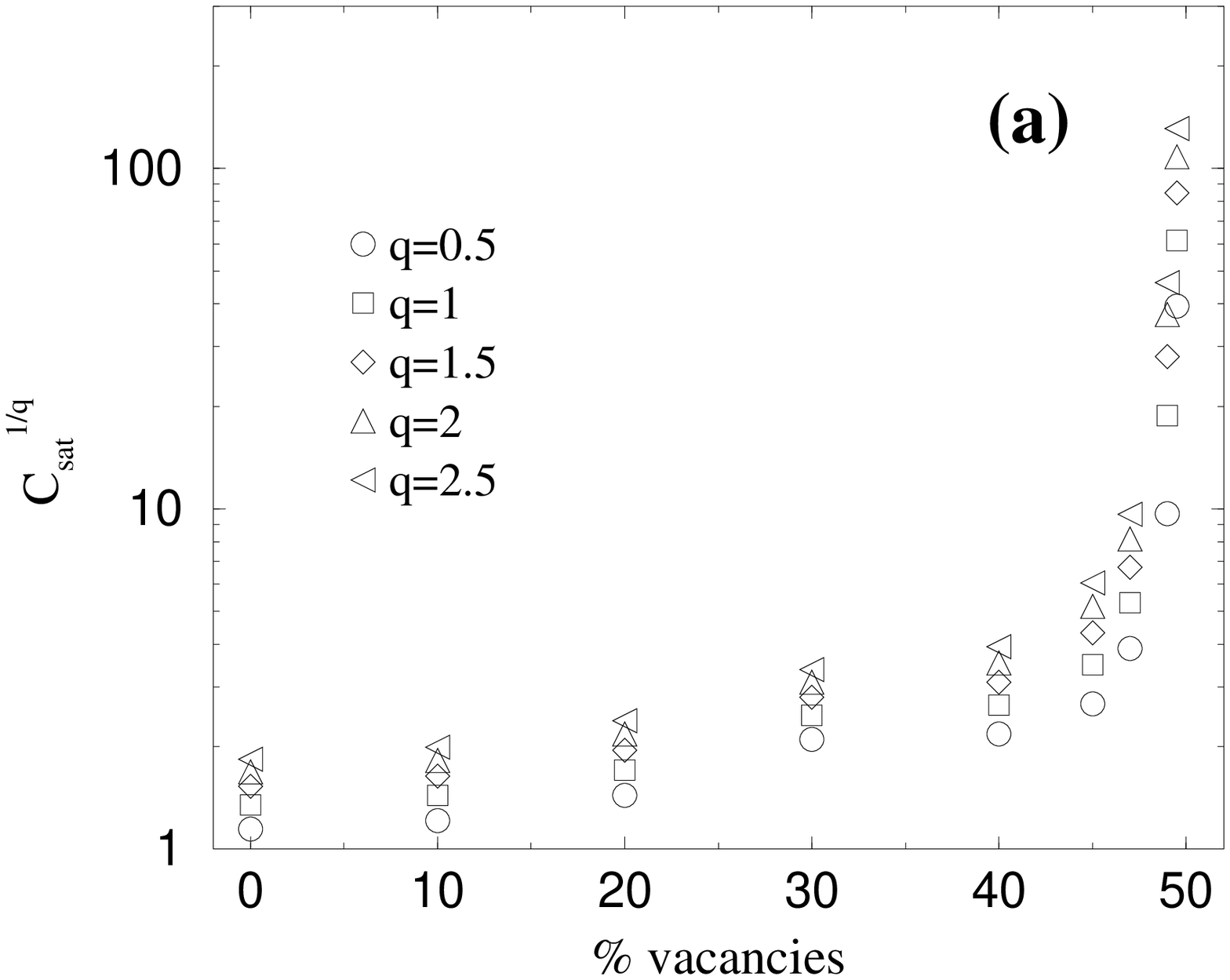}} 
{\epsfxsize=3.4in \epsfbox{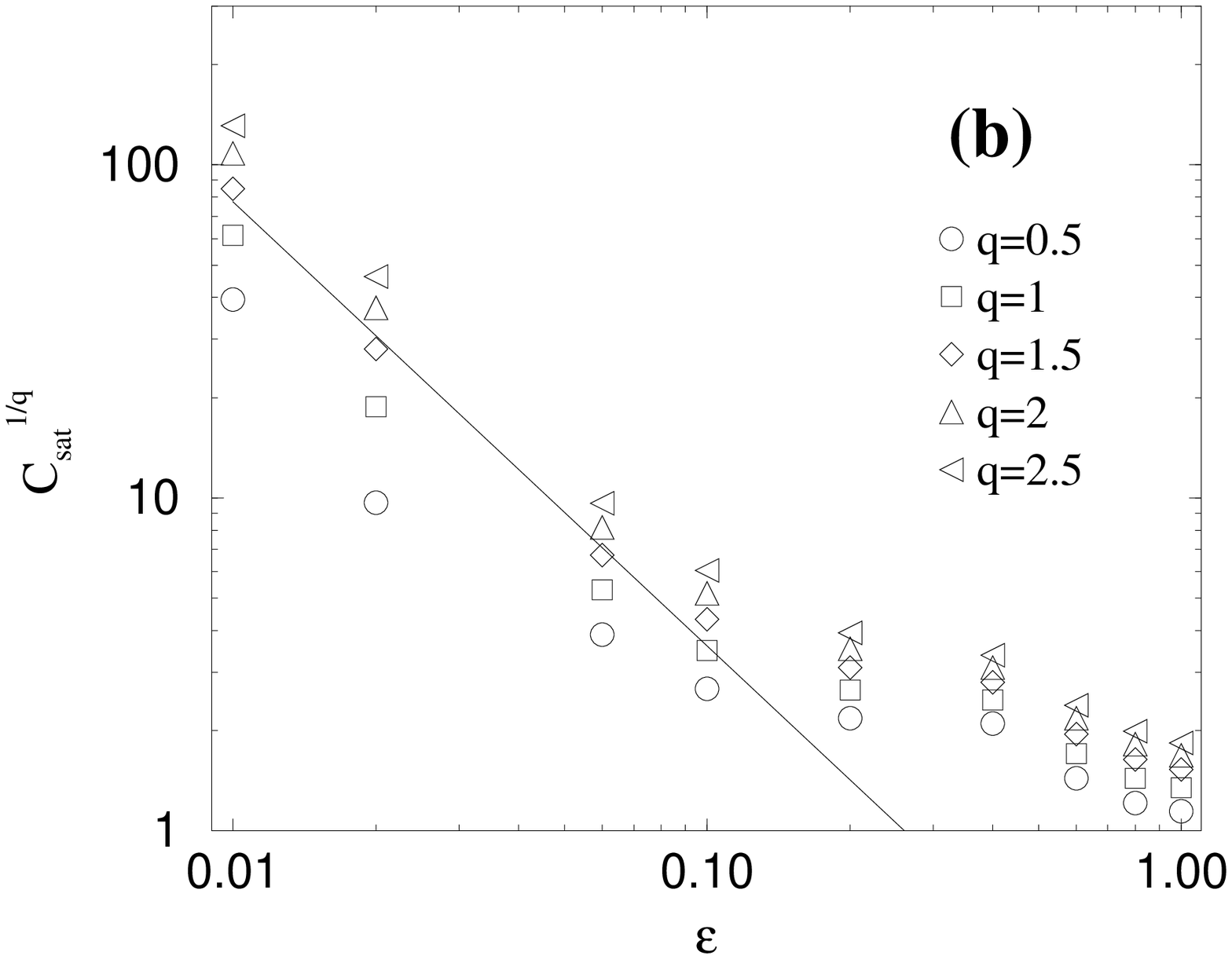}} 
\caption[]{The large-$L$
saturation value of $C_q^{1/q}(L)$, $C_{{\rm sat}}^{1/q}$. 
(a) 
Shown vs vacancy concentration on a linear-log scale. 
(b)
Shown vs $\epsilon = 1 - (\% {\rm vacancies}) / 50\%$ on a log-log scale. 
(Note that the vacancy concentration here increases toward the {\it left\/}.) 
The straight line is proportional to $\epsilon^{-4/3}$, corresponding to
the connectivity length for bulk site percolation \protect\cite{STAU91,SMIR01}. 
In both parts the error bars are smaller than the symbol size. 
}
\label{fig:sat}
\end{figure}

The large-$L$ saturation values, $C_{{\rm sat}}^{1/q}$, are
summarized in Fig.~\ref{fig:sat}. For a fixed concentration of
vacancies, the saturation value increases with increasing $q$. The
saturation value also increases with the vacancy concentration, quite
dramatically so when the concentration approaches the site-percolation 
limit of 50\%. 
Figure~\ref{fig:sat}(a) shows $C^{1/q}_{{\rm sat}}$ vs the vacancy 
concentration on a linear-log scale, while Fig.~\ref{fig:sat}(b) shows the same 
quantity vs the relative distance from the percolation limit, 
$\epsilon = 1 - (\% {\rm vacancies})/50\%$, on a log-log scale. 
The straight line 
in Fig.~\ref{fig:sat}(b) is proportional to $\epsilon^{- \nu}$ with 
$\nu = 4/3$ being the exact value for the connectivity-length exponent 
\cite{STAU91,SMIR01}. The figure indicates that $C^{1/q}_{{\rm sat}}$ 
appears to be coupled to the connectivity length for vacancy percolation in
the bulk, at least for $q \ge 1$. 

\begin{figure}
{\epsfxsize=3.4in \epsfbox{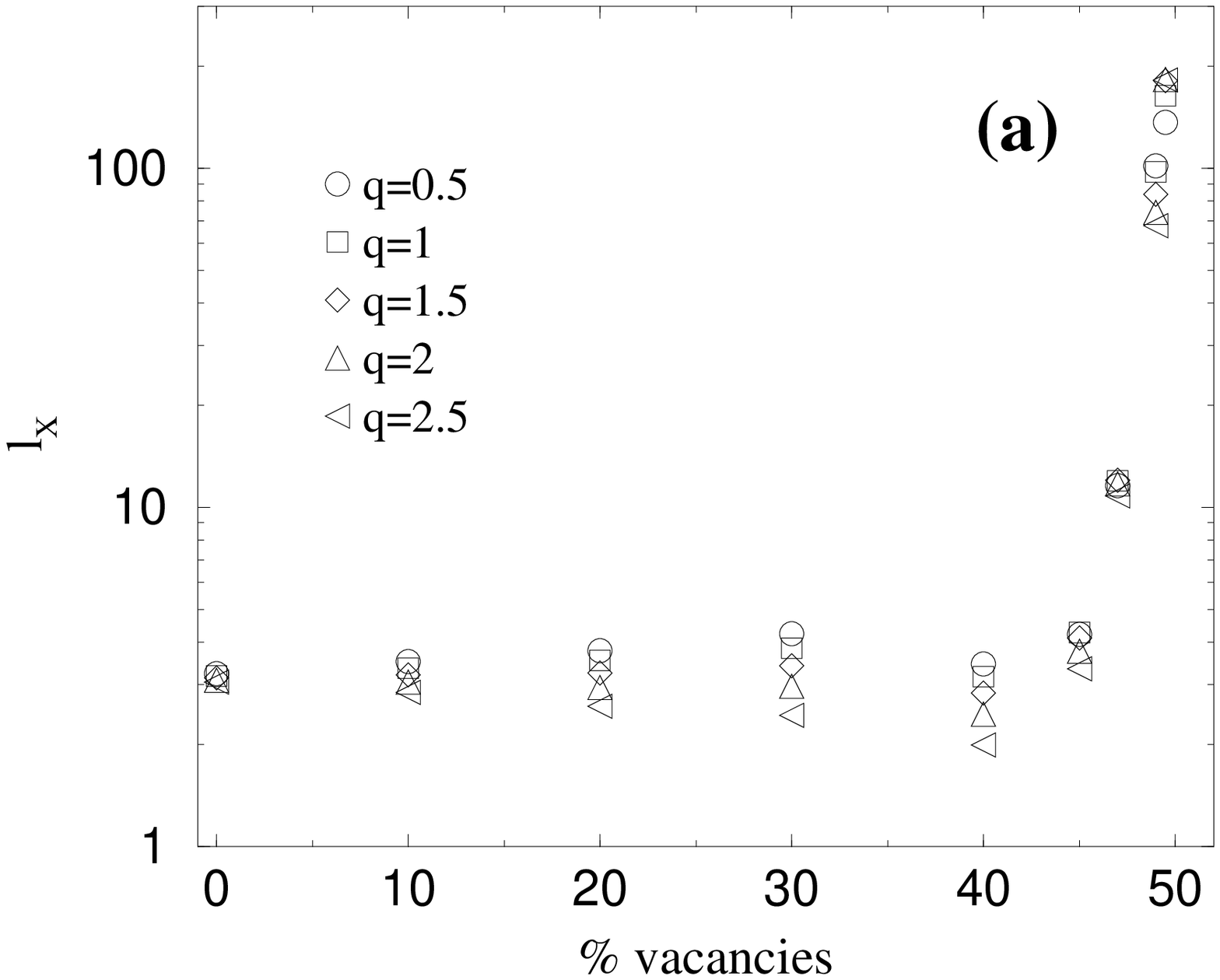}} 
{\epsfxsize=3.4in \epsfbox{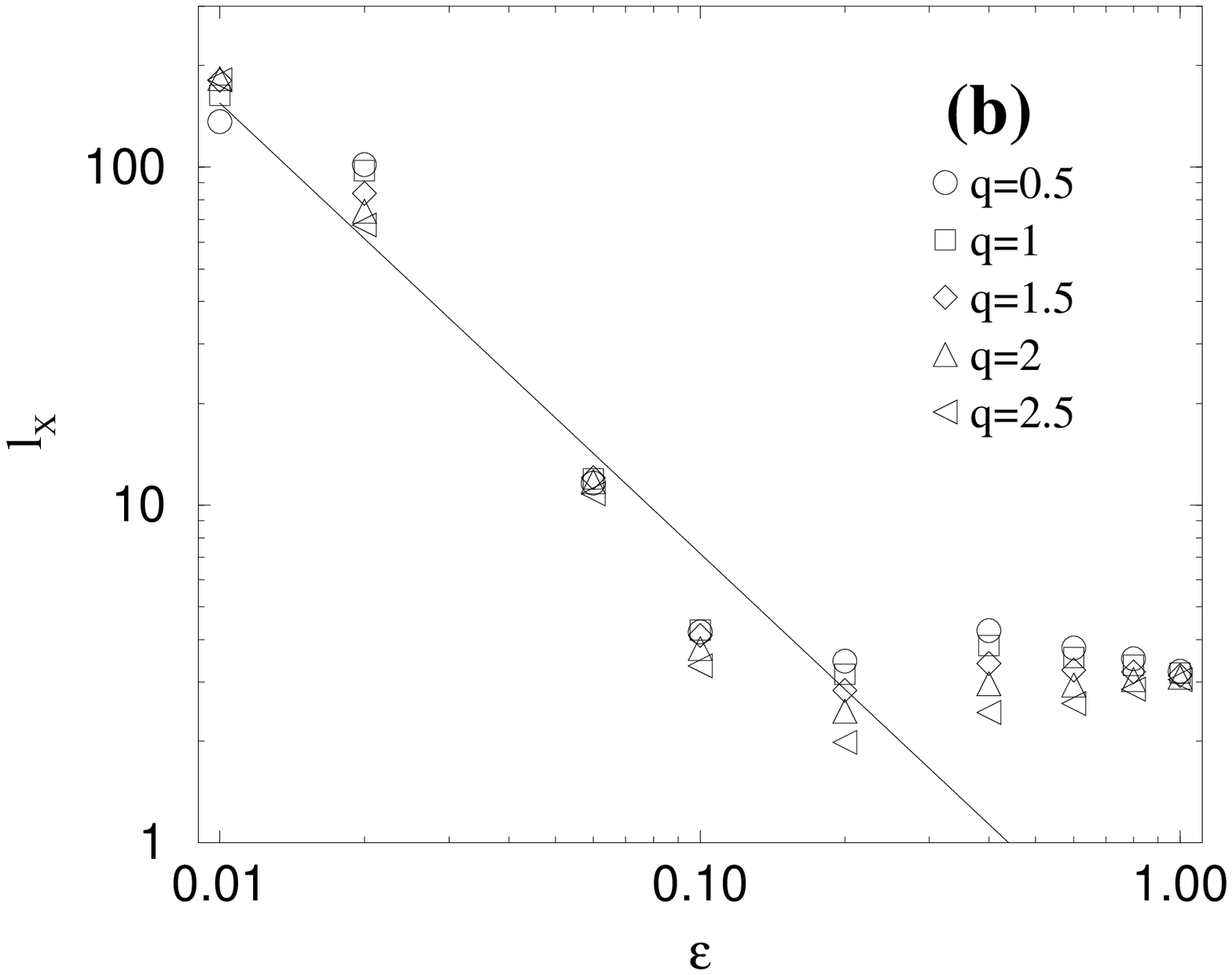}} 
\caption[]{The crossover
length $l_\times$ between the multi-affine regime and the
saturation regime, estimated as the minimum value of $L$ for which
$H_q(L) = 0.3$. 
(a) 
Shown vs vacancy concentration on a linear-log scale. 
(b)
Shown vs $\epsilon = 1 - (\% {\rm vacancies}) / 50\%$ on a log-log scale. 
(Note that the vacancy concentration here increases toward the {\it left\/}.) 
The straight line is proportional to $\epsilon^{-4/3}$, corresponding to
the connectivity length for bulk site percolation \protect\cite{STAU91,SMIR01}. 
In both parts the error bars are smaller than the symbol size. 
}
\label{fig:lcross}
\end{figure}

For a given vacancy concentration, 
all the curves in Fig.~\ref{fig:dq0} coincide for $L \agt l_{\times}$, 
indicating that the surface structure is not scale-dependent on these large 
length scales. Due to the difficulty
in obtaining a precise estimate for $l_{\times}$, we chose to
define it as the value of $L$ at which $H_q=0.3$. 
This value is shown vs the vacancy concentration in
Fig.~\ref{fig:lcross} which, like Fig.~\ref{fig:sat}, is divided into two parts.
In Fig.~\ref{fig:lcross}(a) $l_\times$ is shown vs the vacancy concentration 
on a linear-log scale, 
while in Fig.~\ref{fig:lcross}(b) it is shown vs $\epsilon$ 
on a log-log scale. The increase in the vicinity of the percolation threshold 
is similar to that of $C_{{\rm sat}}^{1/q}$ in Fig.~\ref{fig:sat}, 
except that $l_\times$ 
has a significant background value which dominates at and below 
a vacancy concentration of 40\% ($\epsilon \ge 0.2$). 
Thus our data indicate that both these characteristic surface 
lengths are coupled to the bulk connectivity length as percolation is 
approached. Similar behavior to that observed for the saturation 
value of $C_{2}(L)^{1/2}$ is also seen for the saturation value of $w_L$. 

In the templated polyacrylamide gels studied in 
Ref.~\cite{Mukundan01}, a strong increase in the crossover length and surface 
width 
at surfactant concentrations above 20\% by weight coincides with a change from 
an optically clear material to a white, opaque one. Such a change in the light 
scattering intensity is consistent with an increase in the bulk correlation 
length from much smaller than, to on the order of the wavelength of visible 
light. Considering the numerical and experimental evidence together, we 
conclude that the dramatic increases in the surface-related length 
scales that occur with increasing volume fraction of voids are related to a 
percolation transition in the system of vacancies in the bulk, both in our 
model and in the experimental system by which it was inspired.

\section{Summary and Conclusions}
\label{sec:conc} 
In this paper we have proposed and studied a
simple frustrated spring-network model which, with a relatively modest
computational effort, reproduces aspects of the scaling
behavior of the surfaces of surfactant-templated polyacrylamide
gels, recently observed in AFM experiments \cite{Mukundan01}. 
These results suggest that simple models that incorporate
elastic properties of the gel can provide a useful framework  
for understanding the structural characteristics of these materials, which
are otherwise extremely difficult to simulate due to their complexity. 

The main similarities and differences between the surface structures of the 
model and the experimental system are as follows.
{\it The main similarity\/} is a nontrivial scaling behavior corresponding to 
a power-law form of the increment correlation function at small and intermediate 
length scales. This power-law behavior terminates at a crossover 
length scale, above which the rms
surface width and the increment correlation functions both reach
saturation values that appear to be independent of the system size. 
The origin of this crossover
behavior is explained in terms of the structural inhomogeneities
of the gel. On large length scales the scaling behavior reflects the
average network structure, while on small length scales the
microscopic density fluctuations of the gel network play the most
important role. The quantitative dependence of the crossover on the bulk vacancy 
concentration is discussed further below.
{\it The main difference\/} between the model and the experimental surfaces lies 
in the details of the scaling behavior at length scales below the crossover 
length: the model surfaces are multi-affine, while the experimental surfaces are 
self-affine. The multi-affinity in the model is caused by the finite 
density of vertical discontinuities in the simulated surfaces (see 
Fig.~\ref{fig:zooms}), which are not seen in the experimental system 
(see Fig.~7 of Ref.~\cite{Mukundan01}). The discontinuities in the 
simulated surfaces as they are defined here result from overhangs in the 
spring configuration at the surface, and modification 
of the model to improve the 
detailed agreement with the scaling behavior of the experimental system should 
aim at eliminating these overhangs. 
(It is also possible that overhangs do in fact exist in the undisturbed 
experimental 
surfaces, but are so mechanically weak that they are smoothed out by the force 
of the AFM tip.) 
In the model, the number and size of overhangs could be reduced or eliminated 
in several ways. One possibility is to simulate the smoothing action of the 
AFM tip, {\it e.g.\/}, by a running-average method. Another possibility is to 
include a weak gravitational field (the specific gravity of the polymers is 
greater than that of water), which would bend overhanging sections 
down into the main part of the interface. It should also be noted that 
the larger coordination number in three dimensions most likely would 
reduce the importance of overhangs in a three-dimensional version of the model, 
even without further modifications. (Extension to three dimensions would also 
reduce the number of disconnected pieces, which are partly responsible for 
the large overhangs seen in the case of high vacancy densities. However, the 
computational cost would be large.) A more exhaustive numerical and analytical 
study of the influence of discontinuities on surface scaling behavior is 
currently underway \cite{MICHup}. 

The experimental surfactant templating was simulated by introducing a nonzero
volume fraction of voids in the network through deleting 
lattice nodes and their associated springs. 
Both the horizontal crossover length and the
vertical surface thickness (as measured by the saturation values
of both the rms surface width and the generalized increment correlation
functions) increase dramatically as the vacancy concentration
approaches the site-percolation limit of $50$\%. An analogous
increase of the characteristic horizontal and vertical length
scales with surfactant concentration is observed in the
experimental system, where it coincides with a dramatic increase in 
the optical opacity. The numerical results for our model 
thus support the hypothesis
that this increase in the experimental system is caused by a
percolation transition of the voids created by removal of the surfactants.

In conclusion, 
despite the differences in the details of the scaling behavior at small and 
intermediate length scales, our simplified random-spring model 
successfully mimics several of the surface characteristics of 
an important new class of porous media, including the dramatic increase in 
surface roughness that occurs with increasing surfactant concentration. 
Efforts to further improve the agreement between the small-scale scaling
behavior of our theoretical model and that of the 
experimental system by which it was inspired 
are left for future study. These efforts should in particular involve  
investigations of the dependence of the scaling properties of the model on 
external forces due to gravity or the AFM tip, 
as well as on the statistics of the local spring distribution and the 
dimensionality of the model.

\section*{Acknowledgments}

We acknowledge useful conversations with 
M.~Chakrapani, S.~P.\ Lewis, P.~Meakin, M.~A.\ Novotny, R.~H.\ Swendsen, 
and D.~H.\ Van Winkle. We thank I.~M.\ Navon for bringing to 
our attention the L-BFGS algorithm and Ref.~\cite{subr}, and 
J.~Cardy for telling us about Ref.~\cite{SMIR01}. 
This work was supported in part
by National Science Foundation Grant No.\ DMR-9981815, and by
Florida State University through the Center for Materials Research
and Technology and the School of Computational Science and
Information Technology. 
The computer simulations were performed on the Florida 
State University Physics Department's Beowulf cluster.

\clearpage

\nonumber \noindent
\begin{table}
\caption[]{Scaling exponent $H_q\,$: dependence on $q$ and vacancy
concentration.}
\begin{tabular}{|r|r|r|r|r|r|}
\hline
 Vacancies & $H_{0.5}$ & $H_{1}$ & $H_{1.5}$ & $H_{2}$ & $H_{2.5}$ \\
\hline
0 \% &  1.0008 & 0.9682 & 0.8683 & 0.7030 & 0.5428  \\
\hline
10 \% & 1.0113 & 0.9649 & 0.8232 & 0.6268 & 0.4688  \\
\hline
20 \% & 1.0286 &0.9543  &0.7657 & 0.5627 & 0.42577  \\
\hline
30 \% & 1.0541  &0.9541  & 0.7212 &0.5256  & 0.4088 \\
\hline
40 \% & 1.0910 & 0.9607 & 0.7045 & 0.5168 & 0.4059   \\
\hline
45 \% &1.1087  & 0.9588& 0.6940 & 0.5105 & 0.4019  \\
\hline
47 \% & 1.158 & 0.9713 & 0.6939 & 0.5105 & 0.4019 \\
\hline
49 \% & 1.140 & 0.9662 & 0.6820 & 0.5052 & 0.4021 \\
\hline
49.5 \% & 1.1552& 0.9711  & 0.6772 & 0.5025 & 0.4008 \\
\hline
\end{tabular}
\label{tab:exp}
\end{table}

\clearpage

\clearpage

\clearpage

\newpage

\newpage

\newpage

\newpage

\newpage

\newpage

\newpage

\end{document}